\documentclass[%
reprint,
 amsmath,amssymb,
 aps,
]{revtex4-2}
\usepackage[letterpaper,top=2cm,bottom=2cm,left=3cm,right=3cm,marginparwidth=1.75cm]{geometry}
\usepackage{graphicx}% Include figure files
\usepackage{dcolumn}% Align table columns on decimal point
\usepackage[bottom]{footmisc}
\usepackage{bm}% bold math
\begin{document}

%\preprint{}

\title{Extended Gravity Theories from a Thermodynamic Perspective}
\author{H. R. Fazlollahi}
\email{seyed.hr.fazlollahi@gmail.com (Corresponding Author)}
\affiliation{%
 PPGCOSMO \& Departamento de Física, Universidade Federal do Espírito Santo (UFES), Av. Fernando Ferrari, 514 Campus de Goiabeiras, Vitória, Espírito Santo CEP 29075-910, Brazil}%

\begin{abstract}

We extend the thermodynamic derivation of gravity in the Jacobson framework by generalizing the Clausius relation through a nontrivial entropy functional. We show that entropy deformations universally appear as modifications of the effective gravitational coupling, defining a broad class of modified gravity theories. However, conventional entropy corrections are insufficient to resolve spacetime singularities within this approach. We then propose a new entropy form by incorporating quantum properties at the level of the horizon degrees of freedom. Implementing this entropy within the modified gravitational framework, we investigate its cosmological implications by analyzing both early- and late-time evolution. In the early Universe, the model predicts a nonsingular phase with a finite Hubble parameter, naturally generating a de Sitter–like inflationary expansion with finite entropy and temperature. At late times, the theory reproduces, at leading order, the effective dynamics of loop quantum cosmology.

\begin{description}
\item[Keywords]
Thermodynamics; Gravity Theory; Entropy; Singularity.
\end{description}
\end{abstract}

\maketitle

\section{\label{sec:level1}INTRODUCTION}

The geometrical formulation of gravity introduced by Albert Einstein \cite{Einstein:1915ca} stands as a cornerstone of modern theoretical physics, providing an exceptionally successful description of gravitational phenomena across a wide range of scales. Its predictions have been confirmed with high precision in Solar System tests and astrophysical observations \cite{Einstein:1915bz, Dyson:1920cwa}. Nevertheless, when extended to galactic and cosmological domains, the theory encounters significant challenges \cite{Toomre:1964zx, Berti:2015itd}.

In particular, the observed dynamics of galaxies and large-scale structures suggest the presence of dark matter, while the late-time acceleration of the Universe points toward dark energy \cite{Freese:2008cz, Garrett:2010hd, Peebles:2002gy, Linder:2002et}. Within the standard framework of general relativity, these phenomena require the introduction of additional, yet unobserved, components. At a more fundamental level, the conceptual tension between general relativity and quantum mechanics further indicates that our current understanding of gravity may be incomplete \cite{Kiefer:2004xyv, Rovelli:2004tv}.

These issues have led to two principal research directions. One seeks a unified framework that consistently incorporates both gravity and quantum theory, as in string theory \cite{Gross:1987ar, Witten:1995ex, Boulware:1985wk}. The other explores modifications of the gravitational sector itself, giving rise to a broad class of extended gravity theories \cite{DeFelice:2010aj, Capozziello:2011et, Harko:2011kv, Lovelock:1971yv, Brans:1961sx, Rastall:1972swe, Fazlollahi:2023rhg}. While these approaches have achieved partial success, a fully satisfactory resolution remains elusive.

An alternative and conceptually compelling route is provided by the thermodynamic interpretation of gravity. In a seminal work, Ted Jacobson \cite{Jacobson:1995ab} demonstrated that Einstein’s field equations can be derived from the Clausius relation,

\begin{equation}\label{eq:1}
    \delta Q = T dS,
\end{equation}
applied to local causal horizons. In this framework, gravitational dynamics emerge as an equation of state, with $\delta Q$ representing the energy flux across the horizon and $T$ the associated Unruh temperature. This result strongly suggests that spacetime dynamics may be fundamentally thermodynamic in origin.

Following this insight, considerable effort has been devoted to exploring the interplay between gravitational dynamics and thermodynamic principles \cite{Padmanabhan:2003gd, Eling:2006aw, Padmanabhan:2009vy}. In particular, it has been established that, within a Friedmann–Robertson–Walker spacetime, the Friedmann equations can be reformulated as an expression of the first law of thermodynamics on the apparent horizon \cite{Cai:2005ra, Akbar:2006er}. Such constructions have been extensively investigated in conjunction with generalized entropy functionals, yielding effective phenomenological descriptions of the late-time acceleration of the Universe \cite{Sheykhi:2018dpn, Sheykhi:2022jqq, Fazlollahi:2022bgf, Lymperis:2021qty}. Nevertheless, these approaches are predominantly implemented at the level of cosmological dynamics, typically through modified Friedmann equations rather than emerging from a fundamental revision of the underlying gravitational framework. This limitation naturally prompts a deeper question: whether the thermodynamic perspective can be elevated to the level of the field equations themselves, thereby providing a more fundamental origin for such modifications.

In this work, we examine this possibility by extending the Clausius relation \eqref{eq:1} through the incorporation of quantum corrections in the entropy functional. We demonstrate that a generalized entropy structure, when consistently implemented at the level of the Clausius relation, naturally leads to modified gravitational dynamics, thereby providing a thermodynamically grounded route toward alternative theories of gravity.

The paper is organized as follows. In Sec. II, we briefly review the thermodynamic interpretation of gravity and establish the underlying formalism. In Sec. III, by considering a generalized entropy functional within the Jacobson framework, we derive a broad class of extended gravitational field equations. In Sec. IV, we revisit existing entropy corrections and show that, in this setting, they are insufficient to resolve spacetime singularities. Motivated by this, we introduce a new entropy form incorporating a minimal horizon scale, which ensures a non-vanishing entropy. This formulation is then implemented in the gravitational framework, and its consequences are explored in Sec. V by analyzing the evolution of a spatially flat Friedmann–Robertson–Walker (FRW) Universe, with particular emphasis on singularity resolution. Finally, the main results and conclusions are summarized in Sec. VI.

Throughout this work, we adopt natural units $\hbar = c = 1$. We follow standard conventions in which $dX$ denotes an infinitesimal change of a quantity $X$, while $\delta X$ represents its variation \cite{Jacobson:1995ab}.

\section{Jacobson's Model}

The Einstein field equations can be interpreted not as fundamental dynamical laws, but as a consequence of an underlying local thermodynamic condition imposed on spacetime. In this spirit, Jacobson proposed that, at every spacetime point $p$, the Clausius relation \eqref{eq:1} holds for all local Rindler horizons passing through $p$ \cite{Jacobson:1995ab}.

To formalize this idea, consider an arbitrary point $p$ and construct, in its immediate neighborhood, a local causal horizon generated by a congruence of null geodesics. The generators of this congruence are described by a future-directed null vector field $k^\mu$, affinely parametrized by $\lambda$, such that $k^\mu = \mathrm{d}x^\mu / \mathrm{d}\lambda $. Associated with this horizon is an approximate boost Killing vector field $\chi^\mu$, which vanishes at $p$ and generates local horizon translations. In the vicinity of $p$, this vector field takes the form
\begin{equation}\label{eq:2}
    \chi^\mu = -\kappa \lambda k^\mu,
\end{equation}
where $\kappa$ denotes the surface gravity associated with the local Rindler observer.

Within this framework, the energy flux across the horizon is defined as \cite{Jacobson:1995ab}
\begin{equation}\label{eq:3}
    \delta Q = \int T_{\mu\nu} \, \chi^\mu \, \mathrm{d}\Sigma^\nu,
\end{equation}
where $T_{\mu\nu}$ is the energy--momentum tensor and $\mathrm{d}\Sigma^\nu = k^\nu \, \mathrm{d}\lambda \, \mathrm{d}A$ is the natural surface element on the null hypersurface. Substituting Eq. \eqref{eq:2}, one obtains
\begin{equation}\label{eq:4}
    \delta Q = - \int \kappa \lambda \, T_{\mu\nu} \, k^\mu k^\nu \, \mathrm{d}\lambda \, \mathrm{d}A.
\end{equation}
Since the construction is local, $\kappa$ can be treated as constant over the infinitesimal region, leading to
\begin{equation}\label{eq:5}
    \delta Q = - \kappa \int \lambda \, T_{\mu\nu} \, k^\mu k^\nu \, \mathrm{d}\lambda \, \mathrm{d}A.
\end{equation}
This expression shows that the heat flux is governed by the null projection $T_{\mu\nu} k^\mu k^\nu$, representing the energy flow along the horizon generators.

The temperature is identified with the Unruh temperature,
\begin{equation}\label{eq:6}
    T = \frac{\kappa}{2\pi},
\end{equation}
while the entropy is assumed to satisfy the area law,
\begin{equation}\label{eq:7}
    \mathrm{d}S = \eta \, \delta A,
\end{equation}
with $\eta = 1/(4G)$ and $\delta A$ denoting the variation of the horizon area. The change in area is determined by the expansion scalar $\theta$ of the null congruence,
\begin{equation}\label{eq:8}
    \delta A = \int \theta \, \mathrm{d}\lambda \, \mathrm{d}A,
\end{equation}
where the evolution of $\theta$ is governed by the Raychaudhuri equation \cite{Raychaudhuri:1953yv},
\begin{equation}\label{eq:9}
    \frac{\mathrm{d}\theta}{\mathrm{d}\lambda}
    = -\frac{1}{2}\theta^2 - \sigma_{\mu\nu}\sigma^{\mu\nu}
      - R_{\mu\nu} k^\mu k^\nu,
\end{equation}
where $\sigma_{\mu\nu}$ is the shear tensor. By choosing the congruence such that $\theta = 0$ and $\sigma_{\mu\nu} = 0$ at $p$, one obtains, to leading order in $\lambda$,
\begin{equation}\label{eq:10}
    \theta \simeq - \lambda R_{\mu\nu} k^\mu k^\nu.
\end{equation}
It then follows that
\begin{equation}\label{eq:11}
    \delta A = - \int \lambda R_{\mu\nu} k^\mu k^\nu \, \mathrm{d}\lambda \, \mathrm{d}A,
\end{equation}
and consequently,
\begin{equation}\label{eq:12}
    \mathrm{d}S = - \eta \int \lambda R_{\mu\nu} k^\mu k^\nu \, \mathrm{d}\lambda \, \mathrm{d}A.
\end{equation}

Substituting Eqs. \eqref{eq:5}, \eqref{eq:6}, and \eqref{eq:12} into the Clausius relation yields
\begin{equation}\label{eq:13}
    \int \lambda R_{\mu\nu} k^\mu k^\nu \, \mathrm{d}\lambda \, \mathrm{d}A
    = 8\pi G \int T_{\mu\nu} \lambda k^\mu k^\nu \, \mathrm{d}\lambda \, \mathrm{d}A.
\end{equation}
Since this relation must hold for arbitrary local horizon patches and for all null vectors $k^\mu$ at $p$, the integrands must agree up to a term proportional to the metric tensor. One therefore obtains
\begin{equation}\label{eq:14}
    R_{\mu\nu} - 8\pi G T_{\mu\nu} = \gamma \, g_{\mu\nu},
\end{equation}
where $\gamma$ is a scalar function. Imposing the conservation law $\nabla^\mu T_{\mu\nu} = 0$ together with the contracted Bianchi identity leads to
\begin{equation}\label{eq:15}
    \gamma = \frac{1}{2} R - \Lambda,
\end{equation}
with $\Lambda$ an integration constant. This immediately yields the Einstein field equations,
\begin{equation}\label{eq:16}
    R_{\mu\nu} - \frac{1}{2} R g_{\mu\nu} + \Lambda g_{\mu\nu}
    = 8\pi G T_{\mu\nu}.
\end{equation}
This derivation demonstrates that gravitational field equations can emerge from a purely thermodynamic principle, provided that the Clausius relation holds locally. Crucially, this construction implicitly assumes an equilibrium entropy proportional to the horizon area.

In the following section, we generalize this framework by extending the entropy functional beyond its standard form. By replacing $S$ with a generalized function $f(S)$ in the Clausius relation and following the same thermodynamic procedure, we derive a modified class of gravitational field equations, in which deviations from Einstein gravity arise directly from the extended entropy structure.

\section{Generalized Entropy and Alternative Gravity Theories}

One natural route to extend Jacobson's proposal is to revisit the underlying Clausius relation \eqref{eq:1}. This relation can be generalized in various ways; here, we focus on modifying it through an extension of the entropy functional. In particular, we relax the assumption that entropy depends linearly on the horizon area $A$, and instead allow for a more general functional dependence. A direct inspection of \eqref{eq:7} shows that Jacobson’s original construction implicitly employs the Bekenstein–Hawking entropy,
\begin{equation}\label{eq:17}
    \mathrm{d}S = \eta\, \delta A \quad \Longleftrightarrow \quad S = \eta A,
\end{equation}
However, numerous studies indicate that quantum effects generically induce corrections to this relation, leading to deviations from the standard area law. Such corrections can be systematically encoded by promoting the entropy to a general function of the Bekenstein–Hawking entropy,
\begin{equation}\label{eq:18}
    S_{\text{tot}} = f(S_{\text{BH}}).
\end{equation}
Accordingly, the Clausius relation is modified to
\begin{equation}\label{eq:19}
    \delta Q = f'(S_{\text{BH}})\, T \, \delta A,
\end{equation}
where the prime denotes differentiation with respect to the area $A$. In this way, quantum corrections to the entropy manifest as a deformation of the Clausius relation.

Following the same thermodynamic construction as in Jacobson’s approach, one finds
\begin{equation}\label{eq:20}
    \mathrm{d}S_{\text{tot}} = - f'(S_{\text{BH}}) \int \lambda R_{\mu\nu} k^{\mu} k^{\nu} \, \mathrm{d}\lambda \, \mathrm{d}A.
\end{equation}
Since the analysis is performed locally in an infinitesimal neighborhood around a spacetime point $p$, it is consistent to treat $f'(S_{\text{BH}})$ as approximately constant over the horizon patch,
\begin{equation}\label{eq:21}
    f'(S_{\text{BH}}) \approx \text{const.}
\end{equation}
Thus,
\begin{equation}\label{eq:22}
    \mathrm{d}S_{\text{tot}} = - \int \lambda f'(S_{\text{BH}}) R_{\mu\nu} k^{\mu} k^{\nu} \, \mathrm{d}\lambda \, \mathrm{d}A.
\end{equation}
Substituting this expression into the Clausius relation \eqref{eq:1}, together with Eqs. \eqref{eq:5} and \eqref{eq:6}, yields
\begin{equation}\label{eq:23}
    \int \lambda f'(S_{\text{BH}}) R_{\mu\nu} k^\mu k^\nu \, \mathrm{d}\lambda \,\mathrm{d}A
    = 2\pi \int T_{\mu\nu} \lambda k^\mu k^\nu \, \mathrm{d}\lambda \, \mathrm{d}A.
\end{equation}
Rearranging, and assuming $f'(S_{\text{BH}}) \neq 0$, one obtains
\begin{equation}\label{eq:24}
    \int \left( R_{\mu\nu} - 2\pi \frac{T_{\mu\nu}}{f'(S_{\text{BH}})} \right) \lambda k^\mu k^\nu \, \mathrm{d}\lambda \, \mathrm{d}A = 0,
\end{equation}
which implies
\begin{equation}\label{eq:25}
    R_{\mu\nu} - \frac{2\pi}{f'(S_{\text{BH}})} T_{\mu\nu} = \gamma g_{\mu\nu},
\end{equation}
Here, to determine $\gamma$, we introduce an effective energy–momentum tensor,
\begin{equation}\label{eq:26}
    T_{\mu\nu}^{\text{eff}} = \frac{T_{\mu\nu}}{f'(S_{\text{BH}})},
\end{equation}
and assume its local conservation, $\nabla^\mu T_{\mu\nu}^{\text{eff}} = 0$. Taking the covariant divergence of Eq. \eqref{eq:25} and using the contracted Bianchi identity then leads to the standard result for $\gamma$, Eq. \eqref{eq:15}, yielding the modified gravitational field equations
\begin{equation}\label{eq:27}
    G_{\mu\nu} + \Lambda g_{\mu\nu} = \frac{2\pi}{f'(S_{\text{BH}})} T_{\mu\nu}.
\end{equation}
It is immediately evident that for $f(S_{\text{BH}}) = S_{\text{BH}}$, one recovers the standard Einstein field equations. Within this framework, the entropy generalization manifests as an effective gravitational coupling,
\begin{equation}\label{eq:28}
    G_{\text{eff}} = \frac{1}{f'(S_{\text{BH}})}.
\end{equation}
Thus, quantum-induced deviations from the Bekenstein–Hawking entropy translate directly into modifications of the gravitational interaction strength.

This perspective provides a natural mechanism through which quantum corrections may influence gravitational dynamics, potentially offering new insights into unresolved issues in astrophysics and cosmology. In particular, a well-defined effective coupling could play a significant role in high-curvature regimes, with possible implications for the resolution of spacetime singularities. In the following section, we explore this direction in detail, focusing on entropy models capable of addressing such fundamental limitations.

\section{Entropy corrections}

A wide range of investigations has revisited the validity and universality of the Bekenstein–Hawking entropy \eqref{eq:17} from different theoretical perspectives. In particular, approaches based on quantum information theory, the AdS/CFT correspondence, non-extensive statistical mechanics, the generalized uncertainty principle, loop quantum gravity, string theory, quantum field theory, and quantum entanglement have all suggested modifications to the standard area law \cite{Donnelly:2014fua, Solodukhin:2011gn, Maldacena:1997re, Ryu:2006bv, Tsallis:1987eu, Myung:2006qr, Xiang:2009yq, Meissner:2004ju, Ashtekar:1997yu, Strominger:1996sh}. Despite their diverse origins, these frameworks share a common feature: the entropy is typically expressed as a linear or nonlinear function of the Bekenstein–Hawking entropy, thereby encoding quantum corrections to the microscopic structure of spacetime.

Within some of these fundamental theories, singularity resolution can be achieved at the level of the full dynamical framework. For example, in loop quantum gravity, quantum geometric effects lead to a discrete structure of spacetime, which, in cosmological settings, replaces the classical Big Bang singularity with a quantum bounce \cite{Agullo:2016tjh, Ashtekar:2008zu}. Similarly, in string theory, extended degrees of freedom and duality symmetries modify the ultraviolet behavior of gravity and can soften classical singularities \cite{Barrow:1998wd, Cicoli:2023opf, McAllister:2007bg}. Importantly, these mechanisms arise from the full quantum dynamics of the theory, rather than from entropy corrections alone.

However, within the present thermodynamic framework based on the Jacobson construction, entropy modifications enter exclusively through the Clausius relation and manifest as an effective gravitational coupling, as shown in Eq. \eqref{eq:28}. In cosmological settings, particularly in a spatially flat FRW spacetime, the horizon area shrinks to zero as the scale factor approaches zero. Since conventional entropy corrections depend directly on the area and do not introduce a fundamental lower bound, they fail to regulate divergences in curvature and energy density. Consequently, the initial singularity persists within this class of modified gravity theories.

This motivates the search for an alternative entropy structure capable of incorporating a fundamental cutoff at the microscopic level. To this end, we decompose the total entropy into macroscopic and microscopic contributions,
\begin{equation}\label{eq:29}
    S_{\text{tot}} = S_{\text{mac}} + S_{\text{mic}},
\end{equation}
where $S_{\text{mac}} = \eta A$ represents the Bekenstein–Hawking entropy, and $S_{\text{mic}}$ encodes the contribution of microscopic quantum degrees of freedom associated with the horizon.

To construct $S_{\text{mic}}$, we model the horizon as a two-dimensional surface supporting fundamental excitations. Motivated by the universality of harmonic modes, we assume that these degrees of freedom can be described by a set of independent quantum harmonic oscillators. 

For a single oscillator, the energy spectrum is
\begin{equation}\label{eq:30}
    E_n = \hbar \omega \left(n + \frac{1}{2}\right),
\end{equation}
where $n \in \mathbb{N}_0$. For a system of $D$ independent modes (dimensions), the total energy reads
\begin{equation}\label{eq:31}
    E_{\{n_i\}} = \hbar \omega \sum_{i=1}^{D} \left(n_i + \frac{1}{2}\right).
\end{equation}
Subtracting the ground state energy $E_0 = \frac{D}{2}\hbar\omega$, we define the excitation energy
\begin{equation}\label{eq:32}
    \bar{E} = E_{\{n_i\}} - E_0 = \hbar \omega \sum_{i=1}^{D} n_i \equiv \hbar \omega\, N,
\end{equation}
where
\begin{equation}\label{eq:33}
    N = \sum_{i=1}^{D} n_i
\end{equation}
is the total excitation number. Physically, $N$ counts how many quanta of excitation are distributed among the $D$ independent modes on the horizon. Different sets $\{n_i\}$ can correspond to the same total excitation number $N$, implying that a large number of distinct microscopic configurations give rise to the same macroscopic energy. This degeneracy reflects the underlying statistical nature of the system and plays a central role in the entropy construction.

The total number of such configurations i.e., the number of accessible micro-states is given by
\begin{equation}\label{eq:34}
    \Omega(N) = \binom{N + D - 1}{N} = \frac{(N + D - 1)!}{N!(D - 1)!}.
\end{equation}
This expression counts the number of distinct ways to distribute $N$ identical excitation quanta among $D$ independent harmonic modes. In the regime $N \gg D$, which is appropriate for macroscopic horizons, Stirling’s approximation yields the asymptotic behavior
\begin{equation}\label{eq:35}
    \Omega(N) \simeq \frac{N^{D-1}}{(D-1)!}.
\end{equation}
For a two-dimensional horizon surface, setting $D = 2$ simplifies this result to
\begin{equation}\label{eq:36}
    \Omega(N) \simeq N.
\end{equation}
Using Eq. \eqref{eq:32}, this immediately implies a linear scaling between the number of micro-states and the excitation energy,
\begin{equation}\label{eq:37}
    \Omega \propto \bar{E}.
\end{equation}
At this stage, it is necessary to relate the microscopic excitation energy to the geometric properties of the horizon. In gravitational systems, energy associated with horizons, such as the Misner–Sharp energy or black hole mass, is known to scale with the horizon radius, and therefore with its area \cite{Misner:1964je, Hawking:1975vcx}. More generally, from a holographic perspective, the number of degrees of freedom scales with the area, suggesting that the total excitation energy stored on the horizon should also be an extensive quantity with respect to surface area $A$ \cite{Susskind:1994vu}. Motivated by this, we assume
\begin{equation}\label{eq:38}
    \bar{E} \propto A - A_0.
\end{equation}
Here, $A_0$ plays a crucial role. It represents the minimal area associated with the ground state of the horizon degrees of freedom, corresponding to $N = 0$ or $\bar{E} = 0$. Physically, $A_0$ encodes the existence of a fundamental cutoff: the horizon cannot shrink below this value because the system cannot access energies lower than its ground state. In this sense, $A_0$ introduces a minimal length (or area) scale, which is absent in standard entropy constructions.

Using the statistical definition of entropy \cite{Deutsch:1991msp, Moyal:1949sk},
\begin{equation}\label{eq:39}
    S_{\text{mic}} = \ln \Omega,
\end{equation}
we obtain
\begin{equation}\label{eq:40}
    S_{\text{mic}} = \alpha \ln \left(\frac{A-A_0}{G}\right),
\end{equation}
where $\alpha$ is a dimensionless constant. Finally, the total entropy becomes
\begin{equation}\label{eq:41}
    S_{\text{tot}} = \eta A + \alpha \ln \left(\frac{A-A_0}{G}\right).
\end{equation}
This result resembles the logarithmic corrections encountered in several quantum gravity approaches, but with a crucial distinction: the explicit appearance of the minimal area  $A_0$. Unlike conventional corrections, this structure enforces a lower bound on the horizon size, reflecting the existence of a fundamental ground state for the underlying degrees of freedom.

Nevertheless, this form of entropy introduces a subtle issue. In the limit $A \to A_0$, the logarithmic term diverges, $\ln(A - A_0) \to -\infty$ and consequently, depending on the sign of the parameter $\alpha$, the total entropy behaves as $S_{\text{tot}} \to \pm \infty$. This divergence signals a breakdown of the naive continuum description in the immediate vicinity of the ground-state configuration and indicates that additional physical input is required to properly define the theory in this regime.

There are several natural ways to address this behavior. One possibility is to invoke a discrete structure for the horizon area, motivated by the quantized energy spectrum in Eq. \eqref{eq:30}. In this picture, the area is restricted to discrete values of the form $A = n A_0$, with $n \geq 2$, ensuring that the argument of the logarithm remains finite. Alternatively, one may impose the condition $A > A_0$, thereby excluding the ground-state configuration from the thermodynamic description. In this case, the entropy is well-defined only for excited states of the system, and the singular behavior at $A = A_0$ is interpreted as signaling the boundary of the effective description.

In both perspectives, the theory imposes a nontrivial constraint relating the horizon area $A$ to its minimal value $A_0$, reflecting the underlying quantum structure of spacetime. For instance, adopting the quantized description $A = n A_0$ with $n \geq 2$, and substituting into the total entropy \eqref{eq:41}, one obtains
\begin{equation}\label{eq:42}
    S_{\text{tot}} = n \eta A_0 + \alpha \ln \left[\frac{A_0}{G}(n - 1)\right].
\end{equation}
This expression shows that the entropy remains finite and nonvanishing for all allowed states, while the limit $n \to 1$ is excluded due to the logarithmic divergence. Consequently, the physically accessible configurations correspond to excited states above the ground level set by $ A_0$.

A notable implication of this structure is that the system admits a well-defined minimal configuration characterized by a finite entropy and temperature. The presence of a nonzero lower bound on the horizon area prevents the vanishing of thermodynamic quantities and ensures a regular behavior at the smallest scales. This feature is consistent with quantum mechanical systems, where the ground state is associated with a finite energy and gives rise to nontrivial thermodynamic properties. In this sense, the minimal horizon scale $A_0$ plays a role analogous to the ground state energy, providing a natural regulator for the thermodynamic description.

In the following section, we explore the cosmological implications of this framework by considering a spatially flat FRW Universe in the absence of a cosmological constant, using the modified field equations \eqref{eq:27} together with the entropy form \eqref{eq:41}.

\section{Evolution of Universe}

The determination of restrictions on the quantum number $n$, or more generally on the relation between the horizon area $A$ and its minimal value $A_0$, depends sensitively on the geometry and symmetries of the physical system under consideration. In a cosmological context, a natural arena to investigate these constraints is provided by a homogeneous and isotropic Universe, described by the spatially flat FRW metric.

Within this setting, and in the absence of a cosmological constant, the modified field equations \eqref{eq:27}, together with the entropy form \eqref{eq:41}, lead to the following generalized Friedmann equation:
\begin{equation}\label{eq:43}
    H^2=\frac{8\pi G (A-A_0)\rho}{A-A_0+4\alpha G},
\end{equation}
where $\rho$ denotes the total energy density of the cosmic fluid and $H=\dot{a}/a$ is the Hubble parameter.

For an FRW Universe, the apparent horizon radius is given by $r = H^{-1}$ \cite{Melia:2011fj}, and therefore the corresponding area reads $A = 4\pi H^{-2}$. Substituting this relation into \eqref{eq:43}, one obtains
\begin{equation}\label{eq:44}
    H^2 = \frac{2\pi \left(3 + 2A_0 G \rho \pm \sqrt{\Delta} \right)}{3(A_0 - 4\alpha G)}, 
\end{equation}
with
\begin{equation}\label{eq:45}
    \Delta = (2A_0 G \rho - 3)^2 + 96 \alpha G^2 \rho.
\end{equation}
Requiring consistency with the standard late-time behavior $3H^2 \simeq 8\pi G \rho$ selects the negative branch, yielding
\begin{equation}\label{eq:46}
    H^2 = \frac{2\pi \left(3 + 2A_0 G \rho - \sqrt{\Delta} \right)}{3(A_0 - 4\alpha G)}.
\end{equation}
In the early-Universe limit $\rho \gg 1$, the dominant contribution to $\Delta$ is $\Delta \approx (2A_0 G \rho)^2$, which leads to
\begin{equation}\label{eq:47}
    H^2_{\mathrm{early}} \approx \frac{4\pi}{A_0}.
\end{equation}
Thus, the Hubble parameter approaches a constant value, corresponding to a de Sitter phase \cite{Guth:1980zm, Gibbons:1977mu},
\begin{equation}\label{eq:48}
    a(t) \sim e^{H_{\mathrm{early}} t}.
\end{equation}
This behavior indicates that the model naturally generates an inflationary epoch without introducing additional scalar fields. More importantly, the finiteness of $H$ signals the absence of the standard initial singularity. This regularization is directly controlled by the parameter $A_0$, which acquires a clear geometric interpretation as a minimal horizon area,
\begin{equation}\label{eq:49}
    A_{\mathrm{min}} = A_0 = 4\pi r_{\mathrm{min}}^2 
    \;\longrightarrow\; 
    r_{\mathrm{min}} = \sqrt{\frac{A_0}{4\pi}},
\end{equation}
thereby acting as an effective ultraviolet cutoff.

From Eq. \eqref{eq:47}, the entropy in the early-Universe regime is given by
\begin{equation}\label{eq:50}
    S_{\text{early}}=\frac{A_0}{4G}.
\end{equation}
Comparing this with the total entropy \eqref{eq:41}, one can determine the corresponding value of the horizon area,
\begin{equation}\label{eq:51}
    A= A_0+4\alpha G \mathcal{W}_{0}\left(\frac{1}{4\alpha}\right),
\end{equation}
where $\mathcal{W}_{0}$ denotes the principal branch of the Lambert function. This relation explicitly shows that $A > A_0$, implying that the system never reaches the ground-state configuration. Instead, the minimal area acts as a limiting boundary that cannot be attained dynamically.

In the quantized picture $A = n A_0$, one can derive a consistency relation between the quantum number $n$ and the parameter $\alpha$,
\begin{equation}\label{eq:52}
    4\alpha=\frac{1}{\sigma\, \ln(\sigma)}, \quad\quad \sigma=\frac{G}{A_0(n-1)}.
\end{equation}
Substituting this relation back into the entropy expression \eqref{eq:41} reproduces the result \eqref{eq:50}, demonstrating the internal consistency of the framework. Equation \eqref{eq:52} further implies the constraint
\begin{equation}\label{eq:53}
    n<1+\frac{G}{A_0},
\end{equation}
indicating that the allowed quantum states are bounded and that $n$ cannot grow arbitrarily large. 

The corresponding temperature in the early Universe follows from \eqref{eq:47} as
\begin{equation}\label{eq:54}
    T_{\text{early}}=\frac{1}{\sqrt{\pi A_0}},
\end{equation}
which is manifestly finite. 

In the late-time limit $\rho \to 0$, the Friedmann equation reduces to
\begin{equation}\label{eq:55}
    H^2\approx \frac{8\pi G}{3}\rho-\frac{64\alpha\pi G^3}{9}\rho^2+\mathcal{O}(\rho^3).
\end{equation}
Retaining the leading correction, one obtains
\begin{equation}\label{eq:56}
    H^2=\frac{8\pi G}{3}\rho\left(1-\frac{\rho}{\rho_c}\right),\quad\quad \rho_c=\frac{3}{8\alpha G^2}.
\end{equation}
This expression coincides with the effective Friedmann equation in loop quantum cosmology (LQC) \cite{Agullo:2016tjh, Ashtekar:2008zu}. However, in the present framework, this modification emerges purely from geometric corrections encoded in the entropy functional. Moreover, Eq. \eqref{eq:55} shows that the loop quantum cosmology result corresponds only to the leading-order correction, while the full theory naturally contains higher-order contributions that may become relevant beyond this approximation. It is important to emphasize that, although the present framework exhibits similarities with LQC, the two theories are not equivalent. As demonstrated above, retaining only the leading correction in the late-time regime yields a Friedmann equation of the same functional form as that of LQC. However, this correspondence is limited to the low-energy approximation and does not extend to the full dynamical structure of the theory. A key distinction arises in the early-Universe behavior. In LQC, quantum geometric effects lead to a nonsingular bouncing scenario, replacing the classical Big Bang singularity with a transition from a contracting to an expanding phase. In contrast, the present model predicts a qualitatively different resolution of the initial singularity. As shown in Eq. \eqref{eq:47}, the Hubble parameter approaches a finite constant in the high-density limit, giving rise to a de Sitter–like phase characterized by exponential expansion. In this regime, $H_{\text{early}} \neq 0$, and the cosmological evolution does not involve a bounce, but rather an emergent inflationary phase driven by the underlying thermodynamic structure. Therefore, while the theory reproduces LQC-like behavior at late times, its ultraviolet completion is fundamentally distinct. The transition from a matter-dominated regime to the early-Universe phase is governed by the minimal horizon scale $A_0$, which enforces a finite upper bound on the Hubble parameter and ensures regular cosmological evolution. This highlights that the present construction should be viewed not as a reformulation of LQC, but as a broader thermodynamically motivated framework that encompasses LQC-like corrections as a limiting case while predicting a different high-energy cosmological scenario.

\section{Remarks}

Extending general relativity in a physically consistent manner remains a central challenge in modern theoretical physics. Such extensions are not only motivated by persistent observational tensions from galactic dynamics to cosmic evolution but also by the expectation that gravity should ultimately admit a formulation compatible with quantum principles. However, a large class of modified gravity models is constructed at a phenomenological level, often tailored to address specific observations. While successful in certain regimes, these approaches typically lack a fundamental underpinning, limiting their ability to provide a coherent connection between spacetime dynamics and quantum structure.

In contrast, thermodynamics offers a conceptually robust framework in which gravity can be interpreted as an emergent, macroscopic manifestation of underlying degrees of freedom. Following Jacobson’s proposal, gravitational dynamics can be derived from the Clausius relation applied to local horizons, suggesting that modifications to thermodynamics may provide a principled route toward extending gravity. Given the deep interplay between thermodynamics and quantum theory, such extensions are naturally suited to capture quantum gravitational effects in an effective description.

In this work, we have developed a generalized thermodynamic framework by extending the Clausius relation through a nontrivial entropy functional. Within this construction, all entropy deformations manifest as modifications of the effective gravitational coupling, thereby defining a broad and systematic class of modified gravity theories. An analysis of standard entropy corrections shows that, despite their quantum origin, they are insufficient to resolve spacetime singularities within this framework.

Motivated by this limitation, we have proposed a new entropy form by incorporating quantum properties directly into the horizon structure. The resulting entropy possesses a nonvanishing minimal scale, introducing a fundamental cutoff absent in conventional formulations. Implementing this entropy in the modified gravitational dynamics, we have investigated its cosmological implications. At late times, the theory reproduces, at leading order, the effective dynamics of loop quantum cosmology. In the early-Universe regime, however, it predicts a nonsingular evolution with a finite Hubble parameter, naturally giving rise to a de Sitter–like inflationary phase without invoking additional scalar fields. Notably, both the entropy and temperature remain finite, reflecting the underlying quantum structure encoded in the minimal horizon scale.

The framework presented here establishes a direct link between generalized entropy, effective gravitational coupling, and cosmological dynamics, offering a thermodynamically grounded route toward singularity resolution. At the same time, the present analysis has focused on a specific realization of the theory, and a comprehensive exploration of its implications, ranging from black hole physics to perturbative structure formation, remains to be carried out. We therefore anticipate that further investigation of these entropy-driven modifications may provide deeper insight into the interplay between gravity, thermodynamics, and quantum phenomena.

\section*{ACKNOWLEDGMENT}
The author thanks A. H. Fazlollahi for insightful discussions and for motivating this work. This paper is also dedicated to those far from their homeland, whose memories of home continue to inspire them.

\end{document}